\documentclass[preprint]{vgtc}               

\ifpdf
  \pdfoutput=1\relax                   
  \usepackage{graphicx}                
  \DeclareGraphicsExtensions{.pdf,.png,.jpg,.jpeg} 
\else
  \usepackage{graphicx}                
  \DeclareGraphicsExtensions{.eps}     
\fi%

\graphicspath{{figures/}{pictures/}{images/}{./}} 

\usepackage{microtype}                 
\PassOptionsToPackage{warn}{textcomp}  
\usepackage{textcomp}                  
\usepackage{mathptmx}                  
\usepackage{times}                     
\usepackage{cite}                      
\usepackage{tabu}                      
\usepackage{booktabs}                  

\newcommand{\mycolor}[1]{%
  \begingroup\normalfont
  \includegraphics[height=\fontcharht\font`\B,
  width=20pt]
  {#1}%
  \endgroup
}
\usepackage{todonotes}
\definecolor{mygreen}{RGB}{44,162,95}

\onlineid{1236}

\vgtccategory{Research}

\vgtcinsertpkg

\title{Mapping the Global South: Equal-Area Projections for Choropleth Maps}

\author{Gabriela Molina Le\'{o}n\thanks{e-mail: molina@uni-bremen.de} %
\and Michael Lischka\thanks{e-mail: lischka@uni-bremen.de} %
\and Andreas Breiter\thanks{e-mail: abreiter@uni-bremen.de}}
\affiliation{\scriptsize University of Bremen 
}

\abstract{

Choropleth maps are among the most common visualization techniques used to present geographical data.
These maps require an equal-area projection but there are no clear criteria for selecting one.
We collaborated with 20 social scientists researching on the Global South, interested in using choropleth maps, to investigate their design choices according to their research tasks.
We asked them to design world choropleth maps through a survey, and analyzed their answers both qualitatively and quantitatively.
The results suggest that the design choices of map projection, center, scale, and color scheme, were influenced by their personal research goals and the tasks.
The projection was considered the most important choice and the Equal Earth projection was the most common projection used.
Our study takes the first substantial step on investigating projection choices for world choropleth maps in applied visualization research.

} 

\CCScatlist{
  \CCScatTwelve{Human-centered computing}{Visu\-al\-iza\-tion}{Visu\-al\-iza\-tion application domains};
  \CCScatTwelve{Human-centered computing}{Visu\-al\-iza\-tion}{Visualization techniques}{}
}

\begin{document}


\firstsection{Introduction}

\maketitle



When looking at worlwide phenomena, we often visualize the data on a geographical map to look for spatial patterns.
In the social sciences, choropleth maps are often used to visualize development indicators such as life expectancy 
 and unemployment rate, looking for patterns across world regions. 
%
%
Thus world choropleth maps often appear in data portals of international organizations such as the OECD~\cite{oecd} and the World Bank. 



There are multiple aspects to take into account when designing a choropleth map. 
%
One of them is selecting a projection. Map projections transform the three-dimensional surface of the planet into a two-dimensional plane~\cite{monmonier04}.
However, given the complex shape of the globe, projections can only preserve one of three properties at a time: areas (\emph{equal-area} projections), angles (\emph{conformal}) or distances (\emph{equidistant}). 
Choropleth maps should use an equal-area projection so that each square kilometer has the same size on the screen.
Showing the correct relative areas is an essential property for reliably comparing densities across world regions~\cite{vujakovic19, monmonier04}.
Otherwise, the msp reader may mistakenly interpret the geographical distribution of the encoded variable when comparing the area sizes~\cite{jenny17}.

Certain projections are known to be most appropriate for visualizing specific world regions (e.g., the Albers projection for US maps~\cite{villanueva00}) 
but there are dozens of projections to choose from. Cartographers select according to multiple criteria, including their personal preference~\cite{savric15}.
%
Unfortunately, choropleth maps are often created with 
projections that are not equal-area. This may be due to lack of knowledge 
and to visualization tools like Tableau and Vega-lite~\cite{satyanarayan17} offering the Mercator projection by default.
%

In 2017, Boston Public Schools announced that they would start using world maps with the Gall-Peters projection in their classrooms~\cite{walters17}.
Their goal was to ``decolonize the curriculum'' 
because
the previous Mercator projection exaggerates the size of North America and Europe~\cite{johnson17}.
This change was part of the ongoing debate 
 on how projections may influence the reader's mental image of the world~\cite{monmonier04}. 
Although Peters aimed to empower countries that he felt had been discriminated by cartography, his projection severely distorts the shapes of the tropical regions. Therefore, cartographers have criticized it and suggested using other equal-area projections~\cite{robinson85, savric15} since the organizations that adopt Gall-Peters seem to be unaware of the alternatives~\cite{vujakovic03}.
Monmonier~\cite{monmonier04} however argues that cartographers overrate Mercator as a symbol of Western imperialism.
Having a Eurocentric view of the world could be avoided by simply centering the map on a meridian other than Greenwich.
%

We collaborated with social scientists 
interested in emphasizing the inclusion of the Global South in their field.
\emph{Global South} is a term used by the World Bank referring to low- and middle-income countries, preferred over ``Third World''~\cite{hollington15}.
 It does not refer to the Southern Hemisphere though, as many of those countries are in the Northern Hemisphere and countries like Australia are not included.
%
We conducted a survey with 20 social science researchers to investigate their design choices for choropleth maps according to their research tasks.
We tackled the following research questions:
\begin{description}
\item[RQ1] How do domain tasks influence the design choices for choropleth maps? 
\item[RQ2] What equal-area projection is preferred for world choropleth maps? 
\end{description}

We found that their design choices regarding projection, center, scale, and color, varied according to their tasks.
Overall, the researchers considered the projection the most important choice. They preferred the Equal Earth projection~\cite{savric19} for maps associated with their personal research, and slightly preferred the Gall-Peters projection for a map focused on the Global South. Our study takes a first step on investigating the projection preferences for choropleth maps. 


%




\section{Related Work}
%

Most research on choropleth maps has focused on color choices and pattern detection.
Schiewe's study~\cite{schiewe19} recently confirmed the ``dark-is-more'' bias. 
Multiple approaches have been proposed to compensate their main limitation emphasizing large areas, such as Bayesian surprise maps~\cite{correll17} 
and necklace maps~\cite{speckmann10}.
%
%
%
Recently, Pe\~{n}a Araya et al.~\cite{penaaraya20propagation} compared strategies for analyzing geographical propagation with them.
We focus on other important aspects not yet investigated: its projection and centering in domain-based tasks. 

In 1987, Saarinen~\cite{saarinen87} presented a worldwide study of mental maps of the world. He collected 3863 maps drawn by geography students 
and analyzed how the sketch maps were centered.
He concluded that an Eurocentric image of the world dominated due to colonial influences. 
We include the option of setting the map center to investigate its importance for the domain experts. 
%
Nine years later, Saarinen et al.~\cite{saarinen96} selected a sample of the sketches 
to investigate whether people tend to exaggerate or diminish the size of specific world regions. They found that participants generally exaggerated the size of their home continent and of Europe. Africa was always diminished. 
The researchers argued that the popularity of Mercator caused these mistakes.
Battersby and Motello~\cite{battersby09} later found no evidence that experience with Mercator has any influence on the shape of people's cognitive maps.
We observe that the familiarity with Mercator played a role in the choices of our participants.

The most recommended guide to select a projection is Snyder's manual~\cite{snyder87, jenny17}. It provides a list of projections grouped by region and by property.
Although he suggest over a dozen of equal-area projections for world maps, further criteria to choose one is missing. 
%
%

Projections are not only chosen based on the mentioned characteristics, but also based on personal preferences.
Therefore, a few studies have looked into 
the preferences. 
Werner~\cite{werner93} compared nine projections in a user study with 60 participants. 
%
The most preferred equal-area projection was Eckert IV and the least preferred was Gall-Peters.
%
%
More recently, {\v{S}}avri{\v{c}} et al.~\cite{savric15} conducted a larger study with both cartographers and general map-readers. 
The most preferred equal-area projection was Eckert IV.
However, these studies tested maps for general use with all types of projections. 
We focus on equal-area projections for world choropleth maps. 
We selected projections from both studies and favored the uninterrupted and pseudocylindrical ones according to Werner.
%
As {\v{S}}avri{\v{c}} et al.~\cite{savric15} suggested for future work, we study the preferences based only on the projected landforms. 

Despite its bad performance in the mentioned studies, Gall-Peters is still promoted and adopted by public organizations. 
{\v{S}}avri{\v{c}} et al.\ therefore decided to design a new equal-area projection called Equal Earth~\cite{savric19}, inspired by the well-performing Robinson. 
We include Equal Earth to test whether it would be the most preferred,
and we include Gall-Peters for its association with 
the Global South.
 


\section{Motivation}
We worked together with social science researchers that investigate welfare worldwide.
As welfare research has focused on high-income countries so far, these researchers aim to set the focus on comprehensively including the Global South in their datasets and theories. 
During our collaboration, 
we elicited design requirements and choropleth maps were one of the most common visualizations wished for, to look for geographical patterns~\cite{molina20}. 
When we discussed the first drafts of the world maps, it became clear that using projections different than Mercator and centering the map on somewhere different than Europe were specially important to them, together with having the option of focusing on a region and choosing a color scheme.
%
One reason was that data portals frequently used by the social scientists use the Mercator projection and center the map on the Greenwich meridian.
Going through their data sources, we noticed that the data portals of the World Health Organization (WHO), the World Bank~\cite{worldbank} and Correlates of War~\cite{cow} use the Mercator or the Miller projection for choropleths, neither of them being equal-area. 
%
The portals provided no information about the projections used.



\section{Methods}
%
%
%
%
%

We proposed to the social scientists to co-design the world maps. Therefore, we created a survey 
where they could design their map according to their individual research goals and to the goals of the project they work at.
Our objective was to analyze the choices and the rankings of participants to determine what design parameters are important to them and why, according to the task.


We created a survey with 15 questions and three sections.
The first section was about Task 1, designing a world map according to their personal research goals.
The second section was about Task 2, designing a world map according to the goal of including the Global South.
%
The third section asked demographic information. 
We asked the participants to rate their cartography knowledge on a 5-point Likert scale from ``Very limited'' to ``Very good'' to see whether this influenced their choices.
The survey is included in the supplemental material of this paper.

\subsection{Tasks}
As mentioned above, we prepared two tasks asking the participants to design a world map according to either their personal research goals or the project goal of including the Global South.
%
We created the default maps based on an example by Bostock~\cite{bostock19}. 
The dataset was a standard development indicator, 
life expectancy at birth (2016) from 
the WHO~\cite{who16}.
%
We allowed the participants to customize the map through four parameters: (1) equal-area projection, (2) center, (3) color palette, and (4) geometric scale. 
We randomized the order of the options and did not include the projection names to avoid bias.

Based on the survey conducted by {\v{S}}avri{\v{c}} et al.~\cite{savric15}, the characteristics of the Equal Earth projection~\cite{savric19}, and the reflections of Monmonier~\cite{monmonier04}, we propose the following hypotheses:
\begin{description}
\item[H1] The participants will prefer the Equal Earth projection in both tasks.
\item[H2] The participants will consider the map center the most important design choice in Task 2. 
\end{description}

\paragraph{Color schemes}
We included five colorblind-friendly sequential color schemes (\texttt{Blues} \mycolor{blues}, \texttt{Greens} \mycolor{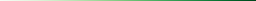}, \texttt{BuPu} \mycolor{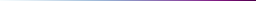}, \texttt{YlGn} \mycolor{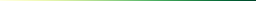}, and \texttt{YlGnBu} \mycolor{YlGnBu}) taken from the well-known color tool, ColorBrewer~\cite{brewer02}. 
All color schemes 
encode higher values with darker colors according to Schiewe's study~\cite{schiewe19}. 
%
We kept the stops of the original color scale to avoid focusing on the scheme choice.

\begin{figure}[]
 \centering
 \includegraphics[width=\columnwidth]{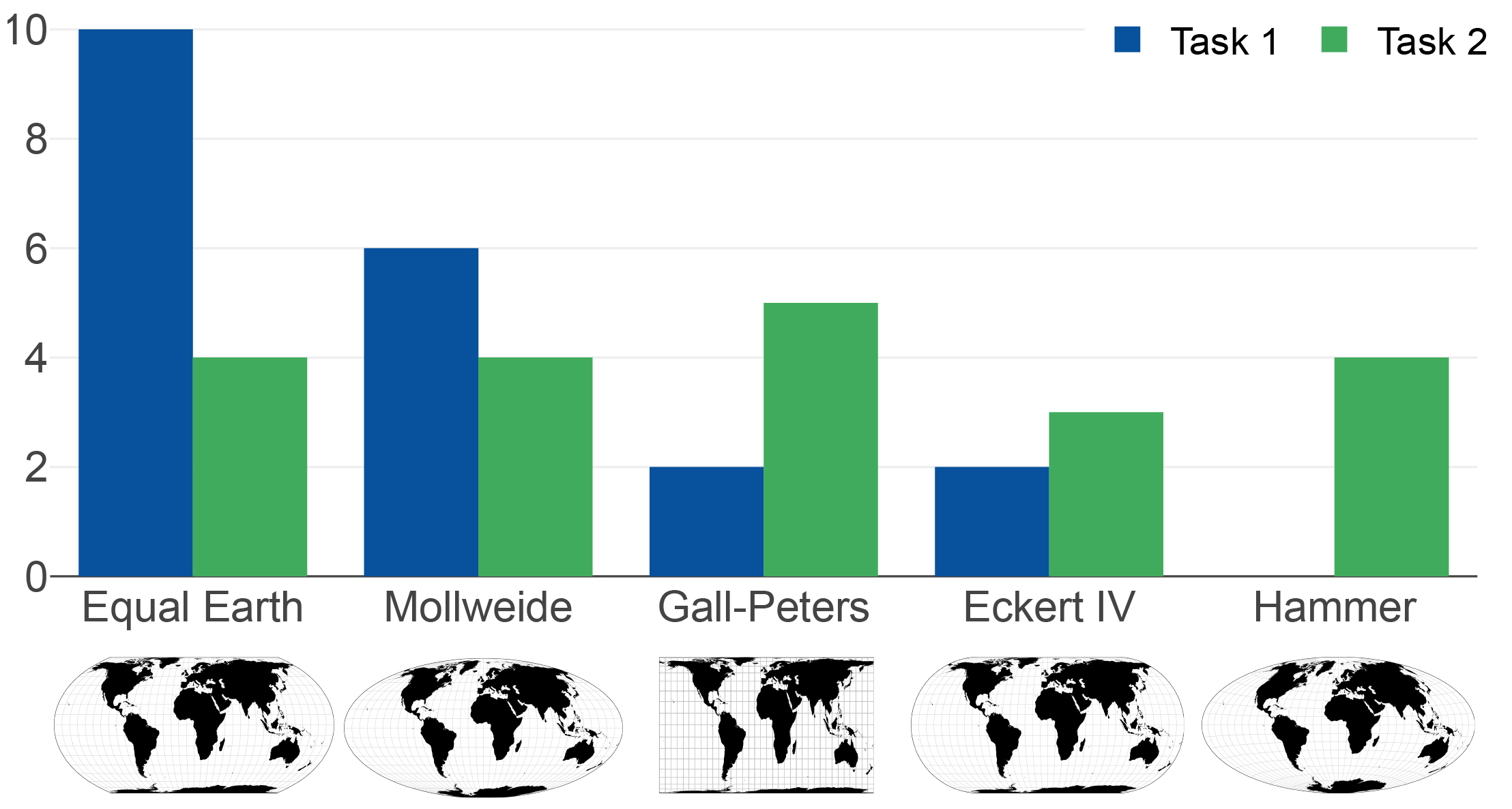}
 \caption{Projection choices for each task. Equal Earth was the most common projection in Task 1 and Gall-Peters was the most common in Task 2. Equal Earth was the most common projection overall.}
 \label{fig:projections}
\end{figure}

\paragraph{Projections}
We included five equal-area projections based on previous preference studies~\cite{savric15, werner93}, recommendations from the literature~\cite{savric19, cairo16},
and projections found in social science portals:
%
%
Eckert IV, 
Mollweide, 
Hammer, 
Equal Earth, and
Gall-Peters. 
The projections are shown in~\autoref{fig:projections} (bottom).
Eckert IV and Mollweide were the most preferred equal-area projections in previous studies~\cite{savric15, werner93}.
Despite its critics, we included Gall-Peters because its conceptual association with ``Third world'' countries can be interpreted as a recommendation for mapping the Global South.
Additionally, we aimed at having diversity regarding the poles representation and the curvature of the parallels. 
The distortion indices of the projections are presented in~\autoref{tab:projections}.
We excluded interrupted projections because those were least preferred in previous studies~\cite{savric15, werner93}. 
We did not reveal the projection names to avoid bias due to the fame of Gall-Peters and the potentially appealing name of Equal Earth.

\begin{table}[]
\caption{Distortion indices and poles representation of the selected projections (from \cite{jenny08, savric14, savric19}). }
\begin{tabular}{@{}llrr@{}}
\toprule
Projection  & Poles as & Scale distortion & Angular deformation \\ \midrule
Eckert IV   & lines                & 0.36             & 28.73               \\
Equal Earth & lines                & 0.37             & 29.08               \\
Mollweide   & points               & 0.39             & 32.28               \\
Hammer      & points               & 0.43             & 35.66               \\
Gall-Peters & lines                & 0.46             & 33.06               \\ \bottomrule
\end{tabular}
\label{tab:projections}
\end{table}

\paragraph{Center}
We included a slider to change the rotation angle $\lambda$ of the projection for setting the horizontal center.
We did not limit the center options to continents or regions because some researchers were interested in multiple countries of different continents.

\paragraph{Scale}
We gave the option of navigating through zooming and panning to adjust what world regions are shown and their size. 

\paragraph{Ranking}
In each task, we asked participants to rank the four design parameters (color, projection, center, and scale) according to their importance for designing the corresponding map. Then, we asked them to shortly explain their reasoning behind the ranking.

\smallskip
We created two computational notebooks for the tasks, using the Observable~\cite{bostock19o} notebook environment.
Using such notebooks allows the domain experts to directly participate in the design process by giving them the time to reflect and freely interact with the prototypes.

\section{Results}
\label{sec:results}

We first had a pilot study and then proceeded to send the survey to the researchers by e-mail, since meeting in person was not possible due to the COVID-19 pandemic.
We conducted the survey with 20 participants (11 female) aged 28 to 69 (mean 36.6). 
%
%
For 15 participants, their home continent was Europe, four came from the Americas, and one from Asia.
Their mean level of knowledge of cartography was $2.15$, being 1 ``Very limited'' and 5 ``Very good''. 


We analyzed the survey responses quantitatively and qualitatively.
We coded the answers based on the grounded theory methodology~\cite{strauss94}.
We applied open coding to the responses and obtained 85 codes. Then we grouped these codes in 13 groups.
This applied to the questions asking for their research goals, their reasoning behind the ranking, and their final comments (if given).
For the quantitative analysis of their projection choices and parameter rankings, we used non-parametric tests since the data did not have a normal distribution.
%
The maps are included in the supplemental material of this paper.

\subsection{Task 1}
\label{sec:results1}

In this task, we asked the social scientists to design a map according to their personal research goals.
These goals covered diverse aspects of welfare or social protection programs: 
six researchers were investigating the development or diffusion of specific social policies worldwide, 
four were interested in the introduction of specific programs,
five investigated health care systems or long-term care labor markets, 
three looked into the benefits for migrant workers, 
and two were interested in the influence of colonialism on inequality and social protection. 
Eight of 20 participants were interested in a specific world region. Half of them were interested in Africa, while others were interested in countries of the Americas, Asia or Europe. 

\paragraph{Projection}
Half of the participants (10) chose the Equal Earth projection for their personal map. Six chose Mollweide. Two chose Eckert IV and two chose Gall-Peters. No one chose Hammer.
An overview of the projection choices for both tasks is shown in~\autoref{fig:projections}. 
P4 chose the Mollweide projection arguing that it seemed the most familiar.
P8 chose Equal Earth because it was ``\textit{aesthetically pleasing}'' but he doubted whether that projection represents correctly area sizes. 
He seemed to believe that a projection could not be aesthetically pleasing and equal-area at the same time. 

\paragraph{Center}
14 participants centered their map on Europe, three on the Pacific Ocean, two on Africa, and one on the Americas.
The participants who had ROIs centered their maps on one of them. 

\paragraph{Scale}
Five participants zoomed in on a region. Four of them zoomed in on their ROI. The fifth person 
had no ROI but wrote that \emph{Scale} was most important for case-oriented research.
These participants ranked it as the first or second most important parameter.

\paragraph{Color}
The most common scheme was \texttt{YlGnBu} \mycolor{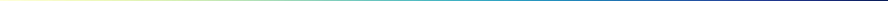}, selected by 11 participants, followed by \texttt{Blues} \mycolor{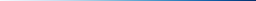} (six participants). 

\begin{figure}[tb]
 \centering 
 \includegraphics[width=\columnwidth]{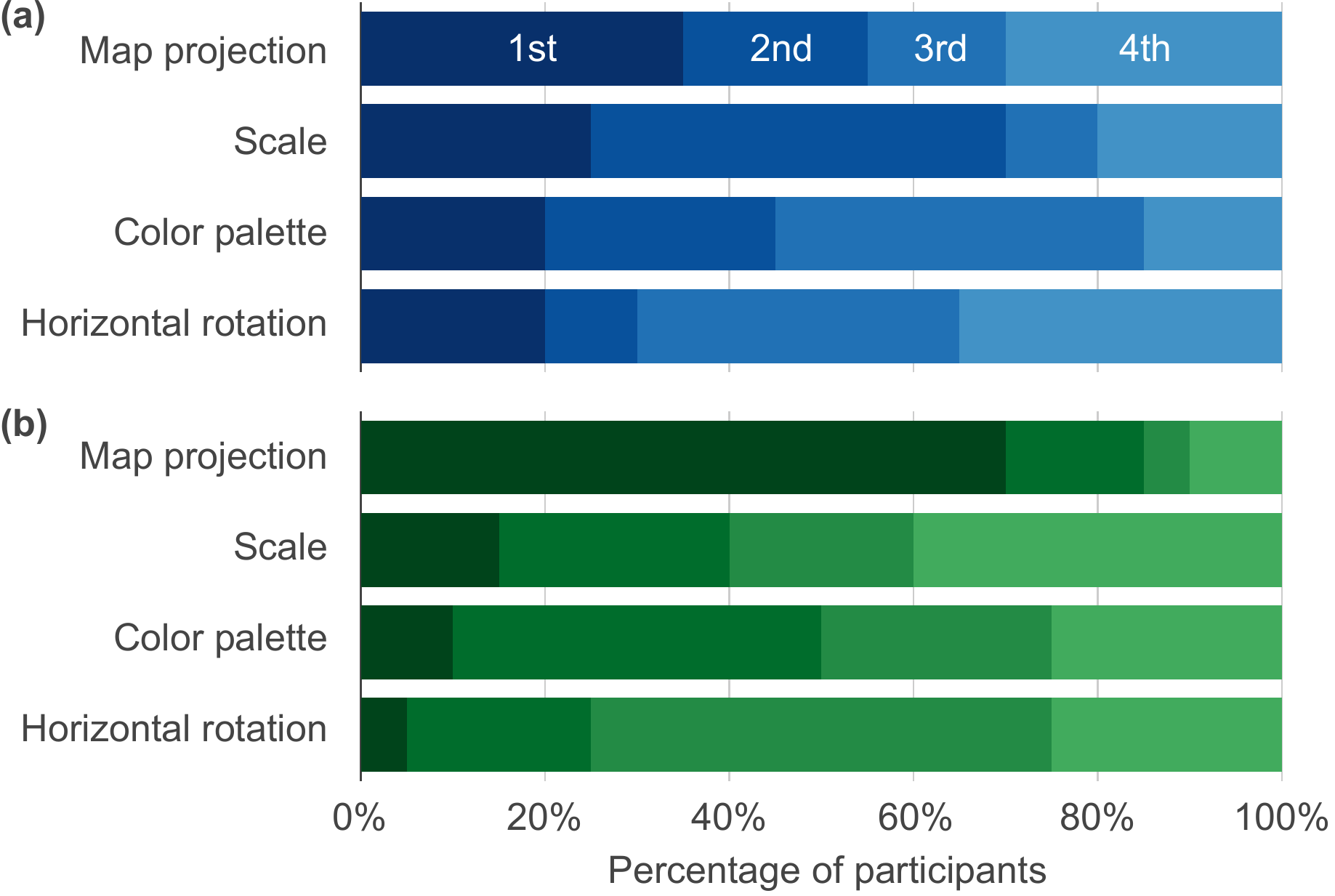}
 \caption{Rankings of the parameters for (a) Task 1 and (b) Task 2.}
 \label{fig:ranking1}
\end{figure}

\paragraph{Ranking}
The map projection was the most important parameter for 35\% of the participants in Task 1. The results are presented in~\autoref{fig:ranking1} (a).
%
We tested the responses to detect any significant patterns on the ranking.
According to the Friedman test, there was no significant effect of the design parameter on the ranked importance for Task 1 ($\chi^2(3) = 2.34, p = 0.505$).

We also tested whether having an ROI affected the rankings. We averaged the rankings of both groups, those who had an ROI and those who did not.
Then we calculated Kendall's $\tau$ to test whether they are correlated. With $\tau =-0.236, p = 0.655$, the result was that there is no strong relationship between the rankings of the researchers with an ROI and the rest.
These results were not unexpected considering that participants had different research goals. 

When asked for their reasoning behind their ranking, the \emph{Map projection} was the most common topic.
%
Four participants that ranked the projection as the most important parameter argued that it helps them to have the best world overview and to distinguish small countries. 

Seven participants that ranked \emph{Scale} as the first or second most important parameter, pointed out that the scale is most relevant for focusing on an ROI. \emph{Center} was often associated with centering on that region. 
Although we did not find a correlation between having an ROI and the rankings, the participants with an ROI set always the center and sometimes the scale, according to them. This then led to different maps.  
%
Two researchers considered the color most important for quickly recognizing the differences between countries. 

%
%

\subsection{Task 2}
In this task, we asked the social scientists to design a map including the Global South.
\paragraph{Projection}
The projection choices are presented in~\autoref{fig:projections}.
In contrast to Task  1, the distribution of the  choices was more equally distributed.
Gall-Peters was the most common projection, chosen by five participants, followed by Equal Earth, Mollweide, and Hammer, chosen by four participants each. 

Seven participants justified their choice based on whether the projection seemed to be equal-area.
They all seemed to have concerns related to the Mercator projection.
Since we did not reveal the projection names, apparently some participants thought that only some options were equal-area.
The seven participants chose all a different option within the five available, and justified their choice as the one seeming most accurate. 
For example, P1 chose Mollweide and wrote that the countries farther from the Equator should not be displayed bigger than they are (a well-known problem of Mercator).
P13 wrote that she had read about Gall-Peters being a good choice, but actually selected Equal Earth. 
She referred to Mercator as ethnocentric.
%
%
%
%

\paragraph{Center}
12 participants centered their map on Europe, three on the Pacific Ocean, and two on the Americas. The other three participants zoomed in on Africa and a neighboring region, centering on the space in between.
The rotation of the map was also associated with the project goals. 
Given that the default center was on the Greenwich meridian, several participants considered that the best way to include the Global South was centering the map on somewhere else than Europe.
P11 argued that since non-OECD countries are spread widely, it was important to have a good overview.
%
In contrast, P2 argued that avoiding OECD-centrism rather depends on the concepts behind the data and not on the map design. 

\paragraph{Scale}
Four participants zoomed in on a southern region. One researcher shifted slightly the focus towards the south while the other three focused on a region with most countries belonging to the Global South.
They all included Africa, two of them focused on the Americas while the other two focused on South Asia. 
They all ranked \emph{Scale} as the first or second most important parameter.
None of them reported having an ROI so there was no connection with it.

\paragraph{Color}
The most common scheme was \texttt{YlGnBu} \mycolor{YlGnBu.png}, selected by 12 participants. Six participants selected \texttt{Blues}, and two \texttt{BuPu}.

\paragraph{Ranking}
The preferences for Task 2 are presented in~\autoref{fig:ranking1} (b).
\emph{Map projection} was the most important parameter for 70\% of the participants, followed by the scale for 15\%.
We analyzed quantitatively the rankings to determine whether some parameters were significantly preferred above others. 
%
According to the Friedman test, there was a significant effect of the design parameters on the ranked importance for this task ($\chi^2(3) = 15, p = 0.002$). 
We performed post-hoc tests for pairwise comparisons to find out what parameters were ranked significantly different.
Nemenyi's test revealed that the \emph{Map projection} was ranked significantly higher than \emph{Color scheme} ($p < 0.01$), \emph{Scale} ($p < 0.01$) and \emph{Center} ($p < 0.05$).
%

Three participants saw the projection as the main tool to avoid Eurocentrism or OECD-centrism by making sure that other countries are represented accurately and the size of high-income countries is not exaggerated. 
%
Eight participants ranked the scale as the least important parameter because their project covers all world regions.

We also tested whether having an ROI affected the rankings in Task 2. We calculated Kendall's $\tau$ to see whether the mean rankings of both groups correlated. 
The result revealed a strong yet not significant correlation between the rankings, with $\tau = 0.913, p = 0.071$.
Participants had similar rankings regardless of their ROIs, possibly due to focusing on the Global South being a common goal.

\subsection{Comparing tasks}

\paragraph{Projection}
The most common projection for Task 1 was Equal Earth and the most common for Task 2 was Gall-Peters. 
Although Equal Earth was most frequently selected overall, 
 the choices differed largely between tasks:
17 of 20 participants chose two different projections for each task. 
The  projection choices for Task 1 were less variant than the choices for Task 2 (see~\autoref{fig:projections}).
%

\paragraph{Center}


12 participants centered both maps roughly on the same region.
Two of them centered their maps on the Pacific Ocean: one wrote that she prefers non-Eurocentric maps while the other researcher gave no reason but her home continent was Asia.
From the eight participants with an ROI, six of them centered both maps on that region.
The map pairs with largest center differences were created by participants who used the center as the main parameter to focus on a region they identified with the Global South. 

\paragraph{Scale}
Most participants who zoomed in wanted to focus either on an ROI or on a Global South region.
We point out that the projection and center choices of the participants who zoomed in may have been different if zooming would not have been possible.
While specific regions look best with certain projections, such projections may distort too much the other regions. 
Nevertheless, we included the scale parameter because it was a design requirement elicited during our collaboration with the researchers. The event logs also suggest that most participants chose a projection before zooming in. 


\paragraph{Color}
\texttt{YlGnBu} \mycolor{YlGnBu.png} was the most common color scheme, used in 23 of 40 maps. 
Eight participants chose a different color scheme for each task. 
Three participants wished for a grey color scheme. 


%

\paragraph{Rankings}
In both tasks, \emph{Map projection} was the highest ranked parameter, although only significantly for Task 2.
%
We compared the rankings for each task per participant by calculating the Spearman's correlation coefficient for each pair of rankings.
The rankings of six participants correlated perfectly ($\rho = 1.0, p < 0.001$) because they ranked the parameters identically in both tasks.
No significant correlation was detected for other pairs.
To measure the correlation between tasks, we calculated the mean ranking per task and compared them. Spearman's results showed that the task mean rankings had no correlation ($\rho = 0.0, p = 1$).

\section{Discussion and conclusions}
Although Equal Earth was the most common projection overall (\textbf{RQ2}), participants preferred Gall-Peters in Task 2. Therefore, we reject \textbf{H1}.
Furthermore, most participants chose two different projections for each task. This suggests that at least the projection choice does differ depending on the task (\textbf{RQ1}).
There is a chance that they changed it because having the option felt like a request to do so.
However, the order of the projections was always random so they could not identify the previous one by their IDs.
Since the projection was the highest ranked parameter in both tasks, we also reject \textbf{H2}. Despite of Monmonier's observation that moving the map center can solve the problem of Eurocentrism~\cite{monmonier04}, our results suggest that projections are considered more important. 

While the choice of map projections has been largely discussed among cartographers, visualization research has not yet looked into criteria for choosing projections for choropleth maps. 
We aim to fill this gap with our study, reporting and analyzing the design choices of domain experts that use these maps in their everyday work.
We conducted a survey with 20 social science researchers to design a world choropleth map according to their research goals.
Our results revealed that choosing a projection is an important design choice, but its importance and selection depend on the task at hand.
The overall preference of the Equal Earth projection suggests that its designers succeeded at their goal of proposing an appealing equal-area projection. 
The preference of Gall-Peters in Task 2 and the reasoning of our participants suggest that this projection is related to the belief that a projection cannot be equal-area and aesthetically appealing at the same time.
%
The different results between Task 1 and Task 2 indicate that the design choices varied according to the task goal. The lack of significant patterns in the maps of Task 1 suggests that no other factors 
consistently affected the design choices. 
The most recognizable pattern was that researchers with an ROI set the center and/or the scale according to the ROI.
%
Further investigating the preferences of general map-readers would help to find out if our findings with domain experts also apply to them.
%
Future work should also consider other map visualizations such as symbol maps. 


\acknowledgments{
We thank the survey participants and the anonymous reviewers for their contributions and feedback. This work was funded by the Deutsche
Forschungsgemeinschaft (DFG, German Research Foundation) –
Projektnummer 374666841 – SFB 1342.
}

\bibliographystyle{abbrv-doi-hyperref-narrow} 

\bibliography{template}
\end{document}